# A Hop-by-hop Cross-layer Congestion Control Scheme for Wireless Sensor Networks

Guowei Wu, Feng Xia*, Lin Yao, Yan Zhang, and Yanwei Zhu
School of Software, Dalian University of Technology, Dalian 116620, China
Email: {wgwdut, fxia, yaolin}@dlut.edu.cn

***Abstract*—Congestions in wireless sensor networks (WSNs) could potentially cause packet loss, throughput impairment and energy waste. To address this issue, a hop-by-hop cross-layer congestion control scheme (HCCC) built on contention-based MAC protocol is proposed in this paper. According to MAC-layer channel information including buffer occupancy ratio and congestion degree of local node, HCCC dynamically adjusts channel access priority in MAC layer and data transmission rate of the node to tackle the problem of congestion. Simulations have been conducted to compare HCCC against closely-related existing schemes. The results show that HCCC exhibits considerable superiority in terms of packets loss ratio, throughput and energy efficiency.**

## I. Introduction

Wireless sensor networks (WSNs) have emerged as an innovative technology that can be applied in a wide range of areas including e.g. environment monitoring, smart spaces, medical systems, and robotic exploration [1]. It has been one of the hot research topics in recent years. The general task of a WSN is to perceive, collect and process information in a cooperative way in the region covered by sensor nodes, and to deliver the information to destination node via certain communication paths. In a sensor node, as data traffic becomes heavier, packets might be put into the node's buffer and have to wait for access to the medium that is shared by a number of communication entities. In such situations, congestion happens in the network. If network congestion becomes severe, certain packets will be dropped due to limited buffer size. This will potentially result in loss of packets, decrease in throughput, and waste of energy. For these reasons, congestion control is a critical challenge facing WSNs [2, 3].

In most cases, conventional congestion control schemes simply reduce the transmission rate at transport layer to relieve network congestion. As a consequence, they cannot maintain stable network throughput. Moreover, existing congestion control algorithms usually do not take into consideration the impact of energy efficiency. The extra signal transmission for the purpose of congestion control and the retransmission of dropped data packets will cause significantly large energy consumption, thus reducing the network life cycle. In addition, in many-to-one multi-hop routing, most of the existing algorithms may cause packets originating from sensors close to the congestion node to have a higher probability of being dropped, which is generally known as the unfairness problem.

To address these shortcomings, a hop-by-hop cross-layer congestion control scheme, namely HCCC, is proposed in this paper. HCCC shares the MAC layer channel information with transport layer and controls network congestion by adjusting data transmission rate and channel access priority. Simulations have been conducted to evaluate the performance of the proposed approach against other related solutions. The results are presented and analyzed.

The rest of this paper is organized as follows. Section 2 reviews related work. Section 3 describes the HCCC algorithm and gives theoretical analysis of its feasibility. Section 4 presents simulation results and performance analysis. Section 5 concludes the paper.

## II. Related Work

A number of congestion control protocols have been proposed for WSNs. The end-to-end and hop-by hop congestion control are two general methods for traffic control in WSNs. The end-to-end control can impose exact rate adjustment at each source node and simplify the design at intermediate nodes. However, it results in slow response and depends highly on the round-trip time (RTT). In contrast, hop-by-hop congestion control has faster response.

CODA (Congestion Detection and Avoidance) [4] is a typical congestion control mechanism in WSNs. It contains three basic strategies: congestion detection based on receiving, open-loop hop-by-hop feedback and multiple source rate adjustment in closed loop. CODA guarantees that throughput satisfies the accurate request by adjusting rate in closed loop way. However, it may cause seriously source rate shaking because of the AIMD (Additive Increase Multiplicative Decrease) strategy, and sensor nodes may deplete extra energy by monitoring channels periodically. ESRT (Event-Sink Reliable Transport) protocol [5] mainly guarantees reliable transmissions and controls congestion by changing and transforming the network state. The SenTCP [6] protocol uses a more accurate method to detect congestion than CODA. It is good for adjusting the sensor node rates properly.

* Corresponding Author: Feng Xia (f.xia@ieee.org)

Cross-layer design can share the information of wireless medium in MAC layer and physical layer with up layers (i.e. network layer, transport layer, and application layer), which can allocate the network resource effectively and hence improve the network performance. Some cross-layer congestion control algorithms have been proposed in recent years. For instance, Hull *et al.* [2] proposed a cross-layer congestion control scheme named Fusion. It exploits three techniques to achieve cross-layer processing: hop-by-hop flow control, rate limiting source traffic when transit traffic is present, and a prioritized MAC protocol. The PCCP [7] algorithm assigns different priorities to every node. It uses cross-layer optimization approach to detect congestion degree and mitigate congestion, where nodes' rates and flows are adjusted according to the priorities. Lin and Shroff [8] presented a cross-layer optimization scheme for multi-hop wireless network, which focuses on how the performance of congestion control will be impacted by imperfect scheduling algorithms. Chiang [9] proposed to jointly optimize congestion control and power control in cross-layer manner. The ANAR [10] mechanism is another cross-layer optimization scheme, which combines transport-layer congestion control and network-layer routing protocol. The Cross-Layer Active Predictive Congestion Control (CL-APCC) scheme [11] for improving the performance of networks applies queuing theory to analyze data flows of a single-node according to its memory status, combined with the analysis of the average occupied memory size of local networks. In order to ensure the fairness and timeliness of the network, the IEEE 802.11 protocol is revised based on waiting time, the number of the node's neighbors and the original priority of data packets. The sending priority of the node is adjusted dynamically. DiffQ [12] provides practical adaptation and implementation of differential backlog that involves a cross-layer optimization of both congestion control and MAC scheduling in real multi-hop wireless networks. ACT (Adaptive Compression-based congestion control Technique) [13] is an adaptive compression scheme for packet reduction in case of congestion. The main problem of ACT is its high complexity. In addition, there are a number of research works attempting to increase the sensor node data transmission throughput, packet delivery ratio and data security via multipath routing [14-18].

Although the existing schemes [3, 19-24] play important roles in improving performance of WSNs, designing an effective congestion control scheme is still a challenging issue in WSNs. In this paper, a hop-by-hop cross-layer congestion control scheme is introduced. The major differences between this work and the aforementioned approaches include the following aspects:

(1) HCCC shares the MAC layer channel information with transport layer, which is used to adjust local channel access probability. When congestion occurs, congestion information can be quickly fed back to upstream nodes, while local congestion can also be alleviated as soon as possible and the local node's buffer queue can avoid being overflowed.

(2) In order to relieve congestion and keep stable network throughput, our HCCC algorithm dynamically adjusts the channel access priority to multiplicatively decrease or linearly increase data transmission rate.

## III. Hop-by-hop Cross-layer Congestion Control

The HCCC algorithm can be built on various contention-based MAC protocols that are widely used in wireless sensor networks. In this work, we adopt the S-MAC protocol [25]. HCCC is composed of three main parts: (1) congestion detection, (2) feedback signal sending and local congestion processing, and (3) feedback signal processing in upstream node. The whole process of the HCCC algorithm is shown in Fig. 1. Suppose that node A, node B and node C are arbitrary intermediate nodes in a WSN, all of which perform the congestion control algorithm. All of them compute their own local congestion information. Node C feeds back its congestion information to its upstream node B. After receiving the feedback signal, node B will add its own congestion information into the feedback signal from node C, and then relay the new feedback signal to its upstream node A. Node B also carries out local congestion processing and feedback signal processing to relieve the congestion within the downstream node C and itself. Node A processes the feedback signal in the same manner with node B, and the feedback signal will be sent to the source node hop-by-hop. Finally the source node will adjust its data transmission rate to relieve congestion. In the following subsections, we will describe the HCCC scheme in detail through elaborating on the used congestion detection method, feedback signal generation and transmission sending method, feedback signal and local congestion processing method.

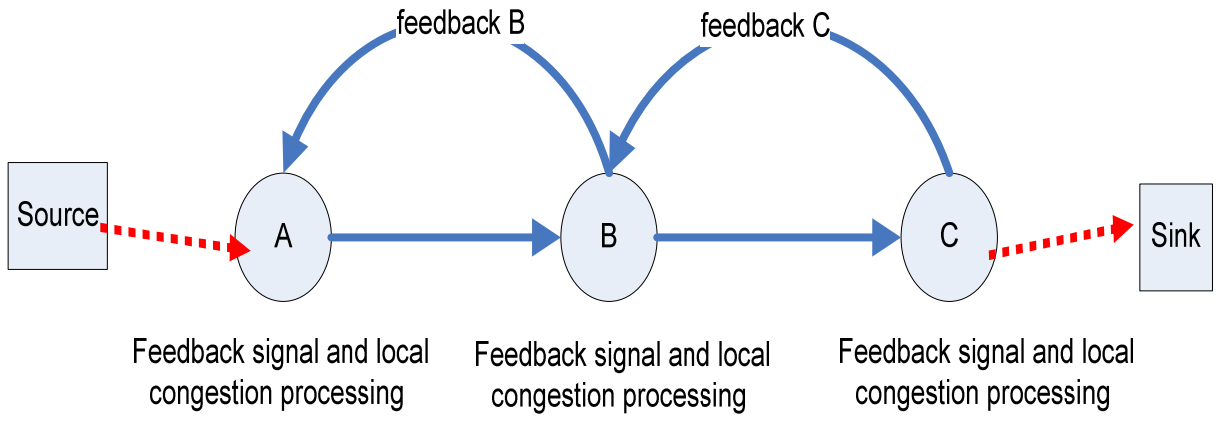

Figure 1. Illustration of proposed scheme

It is worth noting that frequent hop-by-hop transmission of feedback signals would consume significant amount of the node's energy, which is not conducive to prolong network lifetime. To avoid this problem, the HCCC algorithm adopts implicit notice mechanism. The feedback signal is attached in the RTS/CTS control frame of MAC protocol. The MAC layer manages the radio channel and sends the congestion signal to the source node hop-by-hop, which can avoid energy waste caused by broadcast.

### A. Congestion Detection

In order to satisfy the accuracy and low-cost requirements of network congestion detection, HCCC adopts the similar detection mechanism as SenTCP [6] does. We define two parameters: congestion degree $C_d$ and buffer occupancy ratio $B_r$. Congestion degree indicates the changing tendency of buffer queue.

The value of $C_d$ is defined as follows.

$$C_d = T_s / T_a \quad (1)$$

$$T_a = (1-p)\times T_s + p\times(t-t') \quad (2)$$

$$T_s = (1-p)\times(t-t') + p\times t_s \quad (3)$$

In the above equation, $T_a$ is the interval between the arrival of two adjacent data packets in MAC layer, $t$ is the arrival time of the data packet, $t'$ is the arrival time of last data packet, $T_s$ is the average processing time of data packets in local node, $t_s$ is the transmission time of data packets, and $p$ is an adjustable parameter, which is set to 0.3 in this work. The value of $T_s$ is updated when a data packet is sent out. If $C_d > 1$, the arrival rate is bigger than the departure rate of data packets, indicating that congestion may possibly happen in the near future. Let the threshold of buffer occupancy ratio be $B_{max}$. If $B_r > B_{max}$, HCCC can judge that congestion happened. To lower cost, HCCC adopts the most direct way to detect congestion, i.e., using the buffer queue length in local node to detect congestion. When the number of data packets in buffer queue exceeds the threshold value, it is believed that the data packets will overflow the buffer queue in short time, and congestion will happen in the local node. Given below is the congestion detection algorithm.

**Algorithm 1: Congestion Detection**

Input: Node's buffer occupancy ratio $B_r$ and congestion degree $C_d$
Result: Node state change
1: Initialize node information;
2: Compute $B_r$ and $C_d$;
3: if $C_d > 1$ && $B_r > B_{max}$ then
4: Set the node state to congestion state and perform local congestion processing mechanism;
5: Send congestion information (i.e. feedback signal) to upstream node and examine the feedback signal from downstream node;
6: end if
7: if $C_d > 1$ && $B_r < B_{max}$ then
8: Adjust local data transmission rate;
9: end if
10: if $C_d < 1$ && $B_r < B_{max}$ then
11: Set local node state to non-congestion and send node state information to upstream node;
12: end if

As we can see, our congestion detection scheme can not only enable the upstream node to quickly decrease data transmission rate after receiving the congestion signal, but also adaptively adjust local data transmission rate according to congestion tendency.

*B. Feedback Signal Generation and Transmission*

The second phase of HCCC is to generate feedback signal for upstream node and process local congestion. There are three issues that need to be solved in this phase: (1) when to transmit the feedback signal? (2) how to transmit the feedback signal? (3) how to process the congestion locally?

In general, a sensor node may have three states: transmitting, receiving and sleeping. Since a sensor node can only execute the congestion control algorithm when it is not in sleeping state, there are two mechanisms to generate congestion feedback signal: one is to generate feedback signal before the node transmits data packets, and the other is to generate feedback signal before the node receives data packets. Fig.2 and Fig.3 illustrate these two mechanisms respectively, in which node B is the local node, node A is upstream node, and SIFS is the shortest time that physical hardware requires to transform from receiving or detecting state to sending state. The mechanisms are described in more detail in the following.

1) With the first mechanism (Fig. 2), when the local node is ready to transmit data packets, HCCC performs local congestion detection, and adjusts channel access priority according to the congestion condition. The node then sends the congestion information attached in RTS (Request to Send) packet to the upstream node. The upstream node adjusts its channel access priority according to the received congestion information when it begins to transmit a new data packet in next time slot.

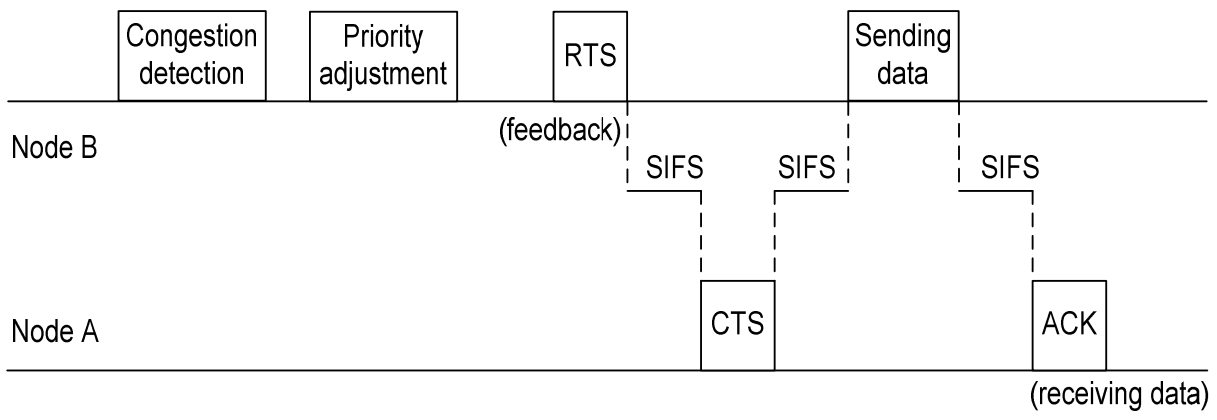

Figure 2. Feedback Signal Generation Method 1

2) With the second mechanism (see Fig. 3), the node performs local congestion detection after receiving the RTS request from the upstream node, then replies to the upstream node with the congestion signal attached in CTS (Clear To Send) packet. The upstream node adjusts channel access priority after receiving the congestion signal when the next data packet is transmitted. The local node adjusts local channel access priority according to the congestion condition when a data packet needs to be sent out.

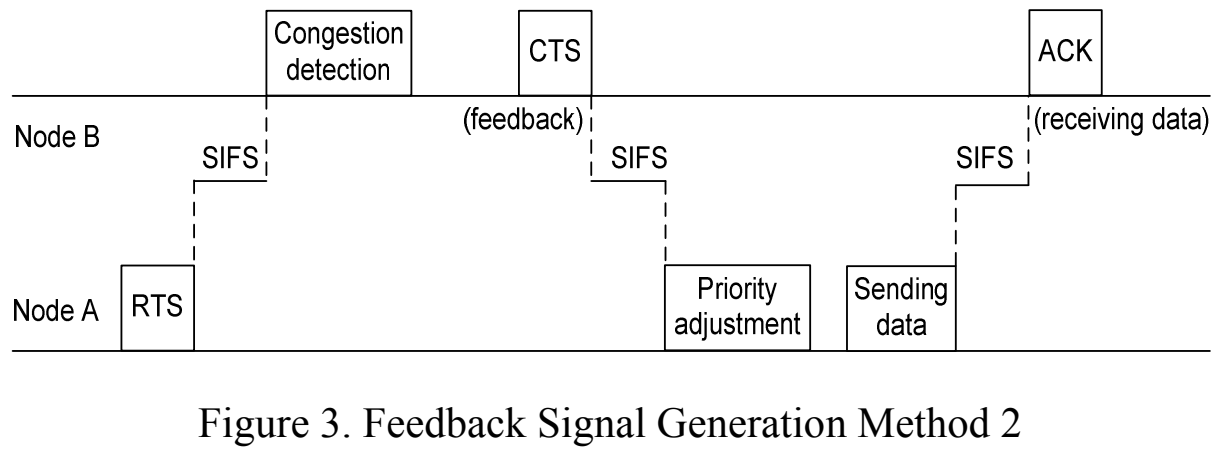

Figure 3. Feedback Signal Generation Method 2

In this work we use the first mechanism for generating and transmitting feedback signals. The reason behind is that it is better to detect congestion and send congestion information before data packets are sent out.

### *C. Feedback Signal and Local Congestion Processing*

The rate adjusting strategy could affect network communication performance significantly, especially network throughput and transmission fairness. When the upstream node processes feedback signals from the downstream node, it should also consider its own congestion condition.

Assuming that the upstream node receives congestion signal successfully, the feedback signal and local congestion processing method used in HCCC is shown in Algorithm 2, where $W$ is the size of local channel competition window, $W_{max}$ is the maximum size of channel contention window in the MAC protocol, and $W_{min}$ is the minimum size of channel contention window. For example, in S-MAC, the range of channel contention window is [1, 63], while it is [15, 1023] in IEEE 802.11. We adopt the same window size as S-MAC. $R$ is local node's data transmission rate. $\Delta R = 0.5 \times (R_{max} - R)$, where $R_{max}$ is the maximum data transmission rate of local node during the period of time from the moment it responds to congestion feedback signal to present. Let $B'_r$ be the buffer occupancy ratio of downstream node, which can be obtained from the feedback signal.

**Algorithm 2: Feedback Signal and Local Congestion Processing**

Input: Local buffer occupancy ratio $B_r$, feedback signal $B'_r$ from downstream node
Result: Local channel contention window $W$, data transmission rate $R$
1: Initialize node information; $W = W_{max}$;
2: if $B'_r > B_{max}$ && $B_r > B_{max}$ then
3: $R \leftarrow 0.25 \times R$; $W = 0.5 \times (5 \times W \times B_r + 0.1 \times W \times 1/B'_r)$;
4: end if
5: if $B'_r > B_{max}$ && $B_r \leq B_{max}$ then
6: $R \leftarrow 0.5 \times R$; $W = 5 \times W \times B'_r$;
7: end if
8: if $B'_r \leq B_{max}$ && $B_r > B_{max}$ then
9: $R \leftarrow \min[0.5 \times R, R + \Delta R]$; $W = \min[10 \times W \times B'_r, 0.1 \times W \times 1/B_r]$;
10: end if
11: if $B'_r \leq B_{max}$ && $B_r < B_{max}$ then
12: $R \leftarrow R + \Delta R$; $W = 10 \times W \times B'_r$;
13: end if

During the hop-by-hop relay process of feedback signals, HCCC gives priority to the handling of local congestion. If $B_r > B_{max}$, which indicates that congestion occurs locally, the node will first send its congestion feedback signal to its upstream node regardless of whether the congestion happens in its downstream node. When $B'_r > B_{max}$ and $B_r \leq B_{max}$, which indicate that congestion happens in the downstream node and not in local node, if the last feedback signal sent out is generated by the local node, then the node will relay the downstream node's feedback signal to its upstream node; otherwise it will not receive the feedback information. This mechanism takes into account both the downstream node's congestion information and the status of the local node. It also avoids frequently relaying congestion information of downstream nodes which would otherwise cause energy waste. When $B'_r \leq B_{max}$ and $B_r > B_{max}$, which indicate that congestion occurs in local node and not in downstream node, the node will only deal with the local congestion without relay. In case of congestion, the local node will reduce its data transmission rate and reduce channel contention window to increase channel access probability. When there is no local congestion, if $B'_r > B_{max}$, indicating that congestion occurs in the downstream node, it will increase channel contention window by $W = 5 \times W \times B'_r$ and reduce the data transmission rate as $R = 0.5 \times R$; otherwise it will linearly increase the data transmission rate as $R = R + \Delta R$.

It is clear that HCCC exploits the AIMD strategy for transmission rate adjustment. The major purpose of using such a strategy is to relieve local congestion as soon as possible while keeping stable network throughput.

### *D. Feasibility Analysis*

**Theorem I:** A sensor node's data transmission rate $R$ is proportional to its channel access priority $P_r$, i.e. $R \propto P_r$

Proof: Suppose that $R^i_{in}$ is the total input traffic rate of node $i$, $R$ is the packet transmission rate at the node $i$ towards node $(i + 1)$, and $R^i_f$ is the packet forwarding rate in the channel, which depends on the channel access priority $P_r$, with high channel access priority implying high probability to access channel and thus yielding high packet forwarding rate. If $R^i_{in}$ is smaller than (or equal to) $R^i_f$, $R$ will be equal to $R^i_{in}$. Otherwise, $R$ will be approximately equal to $R^i_f$. When congestion happens, i.e. $R^i_{in} > R^i_f$, the node's data transmission rate $R$ is approximately equal to $R^i_f$. Since $R^i_f$ is proportional to channel access priority $P_r$, $R$ is proportional to channel access priority $P_r$, i.e., $R \propto P_r$. As a consequence, it is possible to change the transmission rate by adjusting the channel access priority.

**Theorem II:** Channel access priority $P_r$ is inversely proportional to contention window size $W$, i.e. $P_r \propto \frac{1}{W}$.

Proof: Suppose that $S$ is the set of source nodes, $L$ is the set of links, and $c_l$ is the maximum number of packets that can be transmitted in the link in each time slot. Consider a wireless sensor network with $L$ links, each with a fixed capacity of $c_l$, and $S$ source nodes with transmission rate of $R_s$ ($s \in S$). In [8] and [26] it has been obtained that $R_s \propto \frac{1}{D(t)}$, where $D(t)$ is total network delay, including node processing delay, queuing delay and transmission delay. Assume there are $N$ hops from the source to the sink. The queuing delay is random at each hop. Let the value of queuing delay at hop $n$ be $t_{cs,n}$, whose mean value will be determined by the contention window size $W$, and is denoted by $t_{cs}$. It is held that $t_{cs} \propto W$ . The transmission delay will be fixed if the packet length is fixed, which is denoted by $t_{tx}$. Accordingly the entire delay $D(t)$ over $N$ hops is:

$$D(t) = \sum_{n=1}^{N} (t_{cs,n} + t_{tx}) \quad (4)$$

Therefore, we can get $R_s \propto \frac{1}{D(t)} \propto \frac{1}{W}$ , i.e., $R_s$ is inversely proportional to contention window size $W$. Since Theorem I proves that $R$ is proportional to channel access priority $P_r$, channel access priority $P_r$ is inversely proportional to contention window size $W$, i.e., $P_r \propto \frac{1}{W}$ . Consequently, it is feasible to reset the channel access priority through adjusting contention window size.

## IV. Performance Evaluation

To assess the performance of the proposal scheme, we simulate four congestion control schemes including HCCC, CODA [4], ESRT [5] and Fusion [2] using the NS2 simulator. We analyze the performance of these four mechanisms in term of packet loss ratio, throughput, average source transmission rate (i.e. average transmission rate of source nodes) and energy efficiency.

The simulation parameters are set as follows. 100 sensor nodes (including source nodes and sink nodes) are randomly distributed in a square region of 100m×100m. The nodes' communication radius is 30m. The routing protocol used is DSR (Dynamic Source Routing). The network bandwidth is 2Mbps. The transmission rate is 1Mbps. The threshold of buffer occupancy ratio $B_{max}$ is set to 0.4. The range of channel contention window size is [1, 63]. The initial energy of sensor nodes is 0.1J. The main energy consumption of nodes for delivering data packets is $10^{-4}$J/packet. The total buffer size is 500 data packets. The size of every packet is 200 Bytes. The offered load is 5 packets/second (pps).

### A. Packet Loss Ratio

Fig. 4 depicts packet loss ratios (with respect to time) associated with the four congestion control schemes. It can be seen that the packet loss ratio with HCCC is lower than that of CODA and ESRT most of the time. The superiority is especially clear during the beginning stage of system running. HCCC yields almost the same packet loss ratio with Fusion.

### B. Network Throughput

Fig. 5 compares network throughput for these four schemes. The results show that the throughput with HCCC is much higher than the other three schemes, especially when the bit error rate becomes larger. In addition, there is less significant fluctuation when HCCC is used.

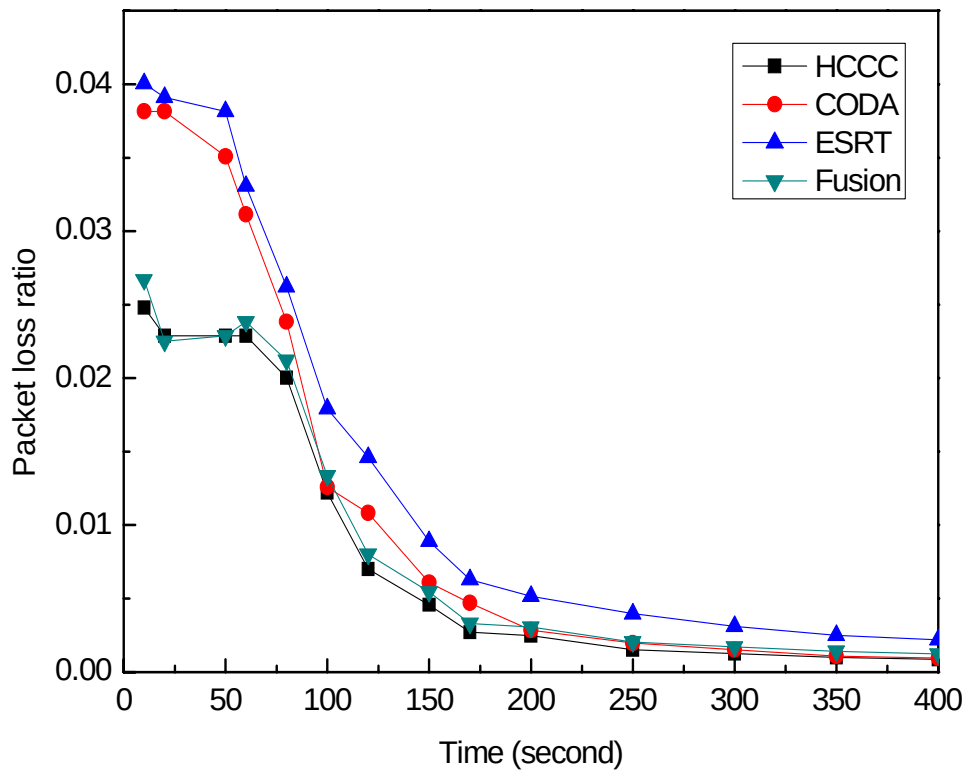

Figure 4. Packet loss ratio

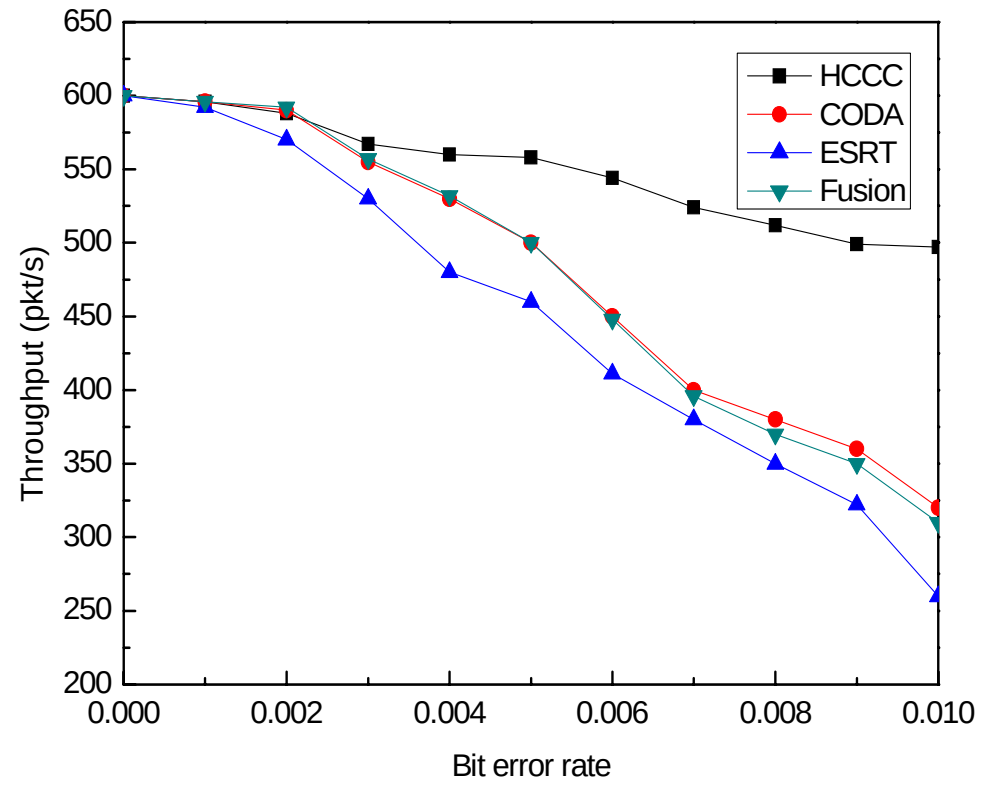

Figure 5. Network throughput

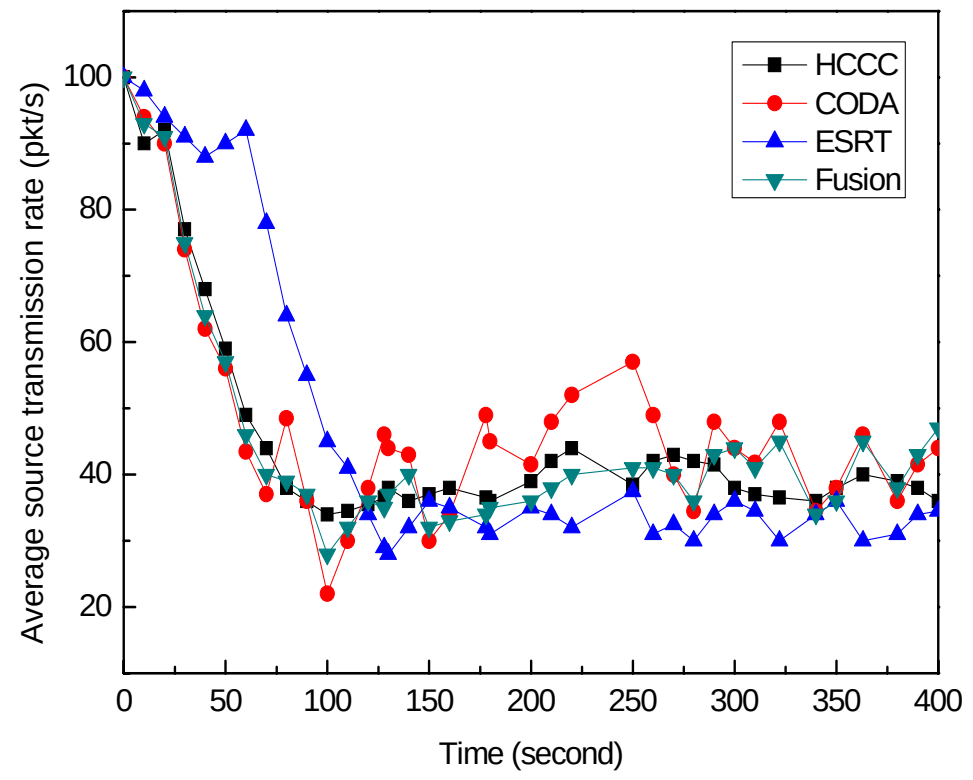

Figure 6. Average transmission rate of source nodes

### C. Source Transmission Rate

Fig. 6 shows the average transmission rate of source nodes under different schemes. The source transmission rate with HCCC maintains in quite a stable level after the transient process. Though the performance of four schemes is comparable in this regard, CODA yields a bit more serious fluctuation in the steady state than the others.

*D. Energy Efficiency*

Fig. 7 gives the comparisons in terms of energy efficiency, which is calculated as the ratio of the sum of nodes' remaining energy to the sum of nodes' initial energy. We can see that HCCC has the highest energy efficiency, while ESRT performs worst in saving energy.

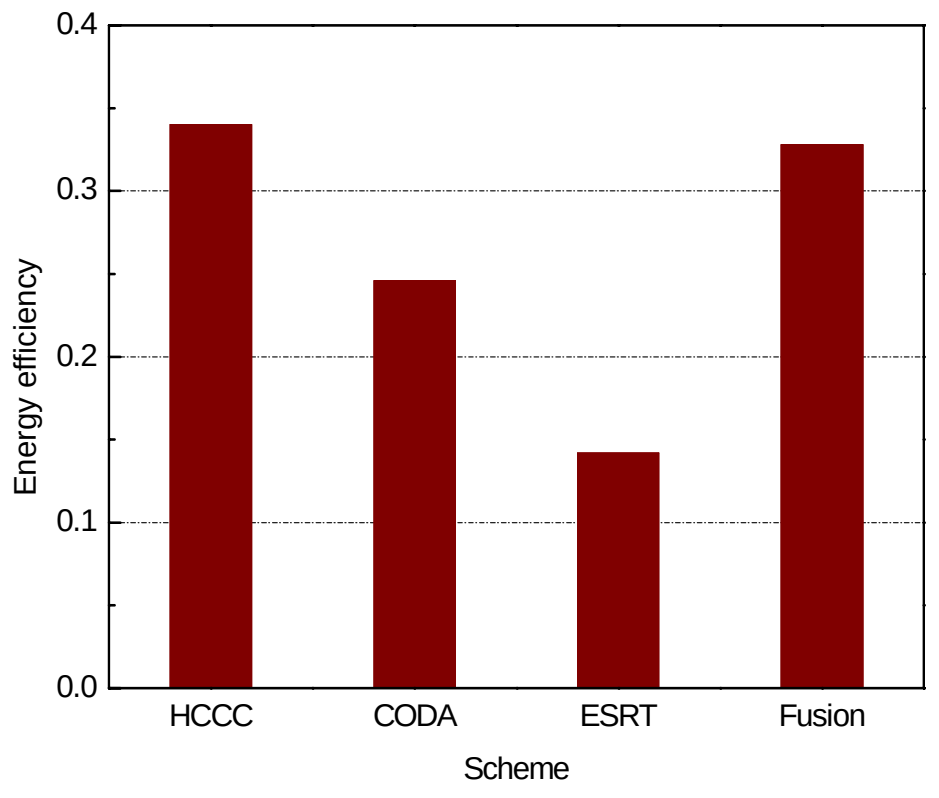

Figure 7. Energy efficiency

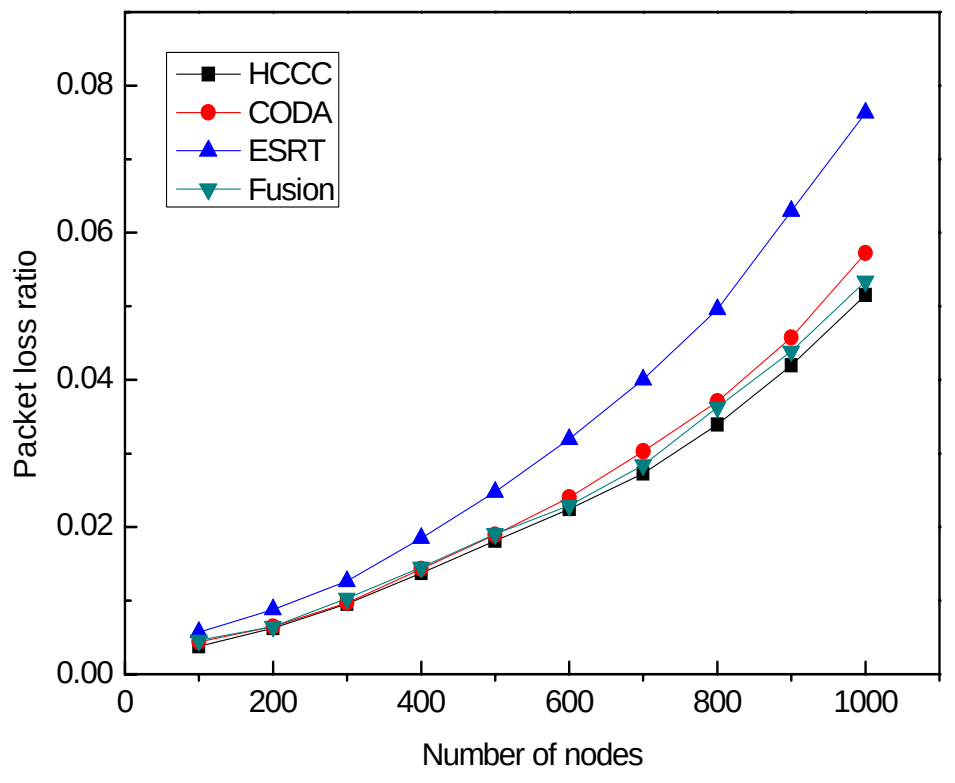

Figure 8. Packet loss ratio with respect to number of nodes

*E. Discussions*

In large scale WSNs, when congestion occurs in one node, the congestion will propagate to other nodes. Congestion regions would consequently form in the network. The more sensor nodes there are, the more packets will be dropped. To address congestion in networks in different scales, congestion control algorithms should have good adaptability. In Fig. 8, we compare the packet loss ratios with different schemes for different numbers of sensor nodes. It is clear that the packet loss ratio increases with the number of sensor nodes. Although HCCC can avoid the propagation of local congestion to its downstream node to some degree, its performance gets worse with the network scale increasing. Therefore, it is not suitable for large scale WSNs.

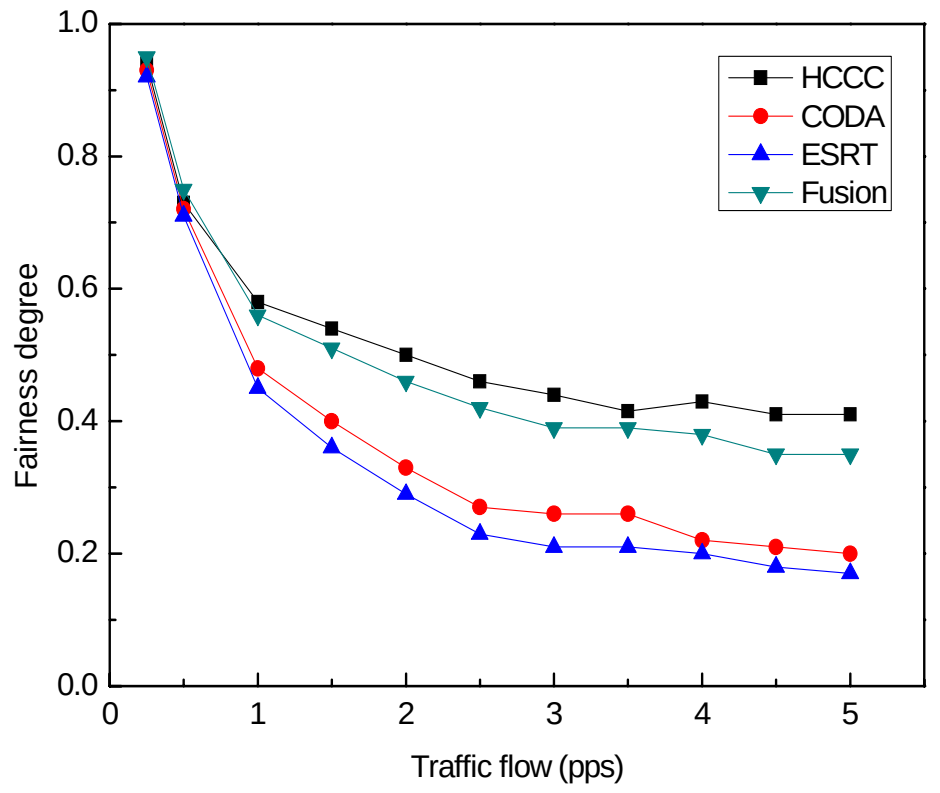

Figure 9. Fairness degree

In Fig. 9, we compare the fairness of the four schemes. In this set of simulations, the fairness degree [27], denoted Φ, is chosen as the metric, which can be computed as:

$$\Phi = \frac{\sum_{i=1}^{N} r_i}{N \sum_{i=1}^{N} r_i^2} \tag{5}$$

where $N$ is the number of nodes and $r_i$ is the average data transmission rate of node $i$. Obviously the value of Φ will vary with data traffic flow. As we can see from Fig. 9, the fairness under the four schemes becomes worse as the traffic flow increases. In particular, CODA and ESRT performs worse in terms of fairness than the others. Although the fairness issue has not been taken into account during the course of HCCC design, it can still yield get relatively better fairness performance than the other three schemes. However, it should be pointed out that there is much room for the improvement of the fairness of HCCC.

## V. Conclusions

In this paper, a hop-by-hop cross-layer congestion control (HCCC) scheme has been presented. HCCC detects local congestion at proper moments, and delivers the congestion information to upstream nodes by exploiting the transmission of RTS and CTS frames. Meanwhile, it adapts the channel access priorities and data transmission rates of sensor nodes. Thus it can adaptively adjust the allocation of channel resource among sensor nodes. The presented simulation results demonstrate that our scheme has good performance in terms of packet loss ratio, throughput, source data transmission rate, and energy efficiency.

## Acknowledgment

This work is partially supported by Natural Science Foundation of China under Grant No. 60903153, the Fundamental Research Funds for the Central Universities, and the SRF for ROCS, SEM.